\documentclass[aps,prb,twocolumn,amsmath,amssymb]{revtex4-1}
\usepackage{amsmath,amssymb,amsfonts,bm}
\usepackage{graphicx}
\usepackage{color}
\usepackage{bbold}
\usepackage{graphicx}
\usepackage{dcolumn}
\usepackage{epstopdf}
\usepackage{color}
\usepackage{epsfig}
\usepackage{url}
\usepackage[pdftex,colorlinks=true]{hyperref}
\hypersetup{citecolor = blue}

\usepackage{verbatim}

\newcommand{\be}{\begin{equation} }
\newcommand{\ee}{\end{equation} }
\newcommand{\ba}{\begin{eqnarray} }
\newcommand{\ea}{\end{eqnarray} }
\newcommand{\n}{\nonumber }
\newcommand{\lf}{\left(}
\newcommand{\ri}{\right)}

\newcommand{\bit}{\begin{itemize}}
\newcommand{\eit}{\end{itemize}}

\begin{document}

\title{Mott, Floquet, and the response of periodically driven Anderson insulators}
\author{Dillon T. Liu$^{1,2}$, J. T. Chalker$^2$, Vedika Khemani$^3$ and S. L. Sondhi$^4$}
\affiliation{$^1$Center for Quantum Phenomena, Department of Physics, New York University, New York, NY, 10003, USA \\ $^2$Theoretical Physics, Oxford University, Parks Road, Oxford OX1 3PU, United Kingdom, \\ $^3$Department of Physics, Harvard University, Cambridge, MA, 02138, USA \\ $^4$Department of Physics, Princeton University, Princeton, NJ 08544, USA}
\date{24 September 2018}

\begin{abstract}
We consider periodically driven Anderson insulators. The short time behavior for weak, monochromatic, uniform electric fields is given by linear response theory and was famously derived by Mott. We go beyond this to consider both long times---which is the physics of Floquet late time states---and strong electric fields. This results in a ``phase diagram'' in the frequency-field strength plane, in which we identify four distinct regimes. These are: a linear response regime dominated by pre-existing Mott resonances, which exists provided Floquet saturation is not reached within a period; a non-linear perturbative regime, which exhibits multiphoton-absorption in response to the field; a near-adiabatic regime, which exhibits a primarily reactive response spread over the entire sample and is insensitive to pre-existing resonances; and finally an enhanced dissipative regime.
\end{abstract}

\maketitle
\section{Introduction}
Many-body localization (MBL) generalizes Anderson localization and entails a breakdown of local thermalization in disordered, interacting systems~\cite{Anderson58, Basko06,  PalHuse,OganesyanHuse, Znidaric,Imbrie2016}. 
Localized systems have been a subject of intense study over the past decade, following a body of work which greatly advanced the case for the existence of MBL using perturbative arguments~\cite{Basko06}, numerical studies~\cite{PalHuse,OganesyanHuse, Znidaric} and rigorous proofs~\cite{Imbrie2016}. MBL systems display a rich complex of properties~\cite{Nandkishore14, AbaninMBLReview} including an emergent set of local integrals of motion\cite{Huse14, Serbyn13cons} leading to a variety of unusual dynamical properties\cite{BardarsonPollmannMoore, Znidaric, nonlocalrearrange, mblcond}. Further stimulus to this study has come from advances in cold atomic systems~\cite{BlochRMP, Choi2016,Smith2015,BlochFloquet, KaufmanEntanglement,GreinerMBLEntanglement2018} which, unlike solids containing delocalized phonons, realize isolated systems in which \emph{all} degrees of freedom are localized --- thus allowing the simplest theory, already quite complicated, to confront experiments directly.

\begin{figure*}
    \begin{tabular}{ccc}
    \raisebox{0.98\height}{ \begin{tabular}{c}
    \includegraphics[height=17mm]{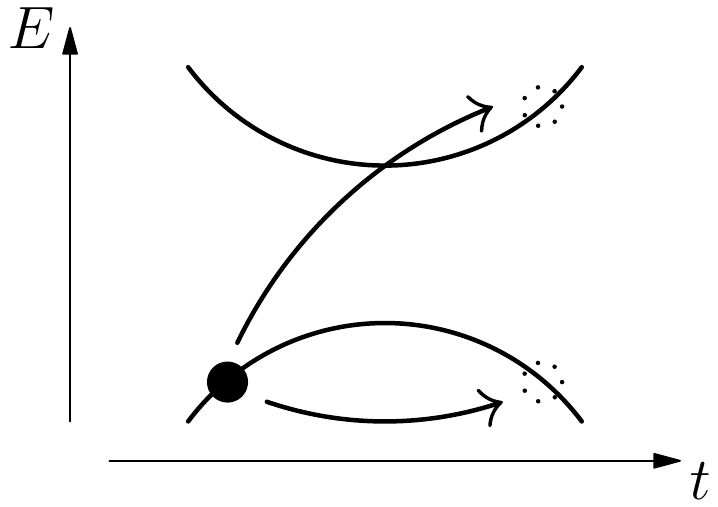} \\
\includegraphics[height=17mm]{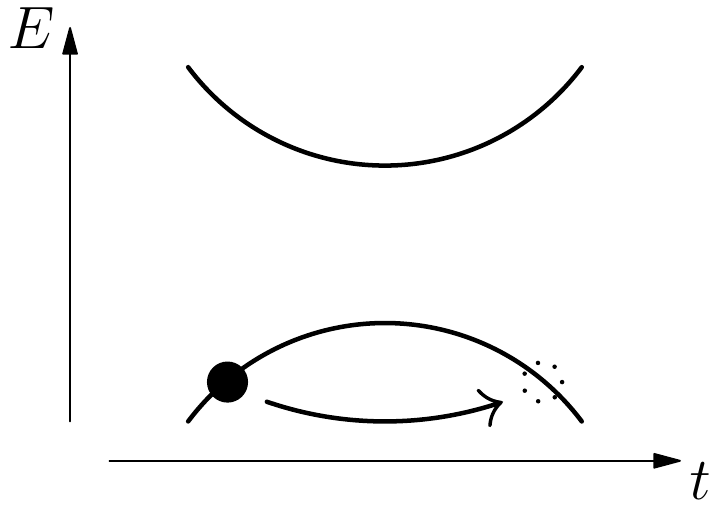} \\
    \includegraphics[width=25mm]{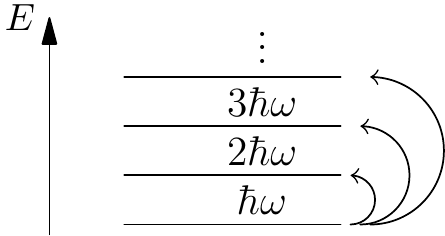} \\
\includegraphics[height=11mm]{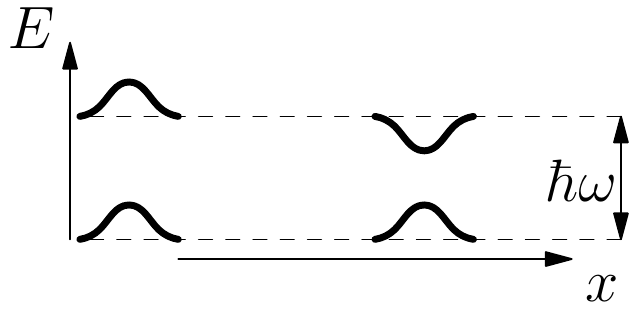}
    \end{tabular}} &
    \includegraphics[width = 0.35\textwidth]{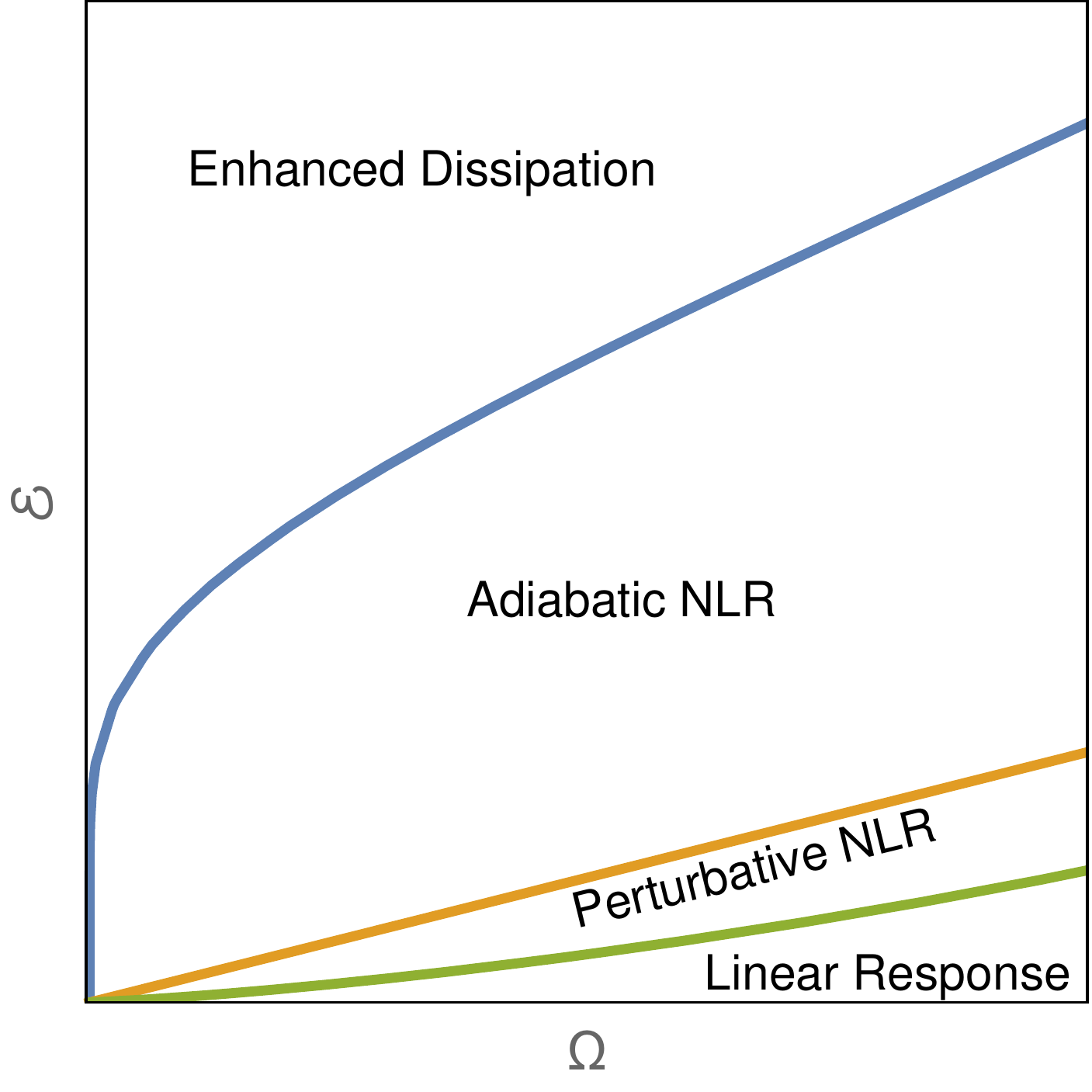} &
    \raisebox{-0.045\height}{\includegraphics[width=0.43\textwidth, height = 0.42\textwidth]{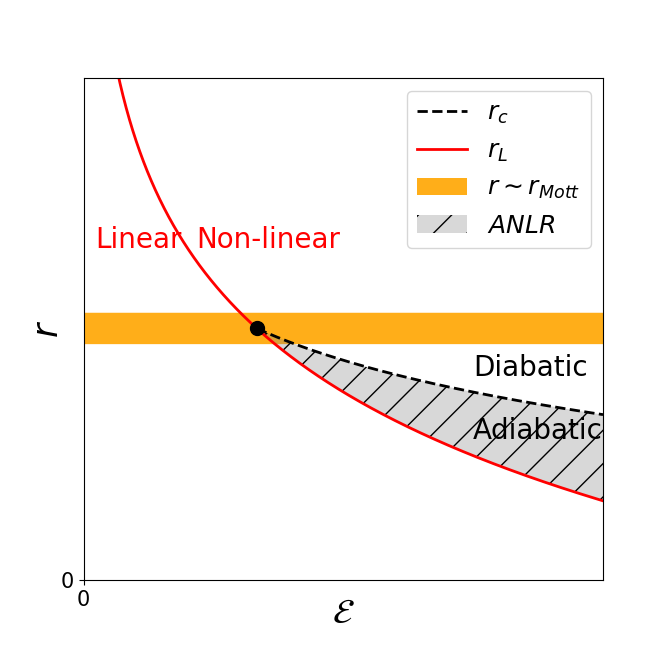}}
    \end{tabular}
    \vspace{-5mm}
    \caption{(left) Schematic illustration of the characteristic microscopic response to an oscillating electric field in the four different regimes. Sketches represent physical processes dominating in the four regimes of response. From bottom to top: Mott resonances, corresponding to the linear response regime; multiphoton absorption, corresponding to the perturbative non-linear response regime; adiabatic avoided level crossings, corresponding to the adiabatic non-linear response regime; and level crossings intermediate between adiabatic and diabatic, corresponding to the enhanced dissipation regime. (middle) ``Phase diagram'' showing four regimes of response for an Anderson insulator driven by an oscillating electric field, as a function of scaled field strength ${\cal E}$ and scaled frequency $\Omega$. See main text for distinctions between regimes. (right) Schematic plot showing the length scales $r_{\rm Mott}$, $r_L$ and $r_c$ that characterize the response of a pair of localized states in an Anderson insulator to a periodic drive, and their dependence on drive strength ${\cal E}$. Linear response is mainly from resonant pairs with separation $r_{\rm Mott}$ indicated by a horizontal band. Other aspects of response depend on the separation $r$ of the localization centers of the pair of states compared to $r_L$ and $r_c$. For $r\ll r_L$ the effect of the drive is perturbative in $\cal E$. For $r\gg r_L$ the pair of states undergoes two avoided level crossings during the drive cycle, which are adiabatic if $r\ll r_c$ and diabatic if $r\gg r_c$. The shaded region with hatching indicates the crossings which dominate the dynamics of the adiabatic non-linear regime.}
    \label{fig:regimesGD_lengthscales}
\end{figure*}

The present paper is inspired by this harmonic convergence, although it addresses non-interacting or Anderson localized systems for reasons of tractability. Specifically we ask about the response of an isolated one dimensional Anderson insulator composed of a single set of charges, which we take to be electrons, initially in its ground state, when it is placed in a uniform electric field oscillating at a frequency $\omega$ and amplitude $E_0$. The textbook answer to this problem is that the system will exhibit a linear response of the celebrated Mott form for the a.c.~conductivity \cite{mott,berezinskii,ivanov,gruzberg} at small $\omega$
\begin{align}\label{MottLawEqn}
\sigma\lf\omega\ri \sim \omega^2 \log^{d+1}\lf 1/\omega \ri.
\end{align}
In this work, we go beyond this answer in two ways. First, we ask what happens when the field is kept on for a long time. Here the linear response calculation, which predicts a linear absorption of energy with time, will break down. Instead we find that the energy absorbed saturates and the system exhibits a Floquet late time state (FLTS). Second, we ask what happens if the field is too large for the linear response formula to hold even at short times. By definition this also involves a breakdown of linear response theory due to the inherent non-linearity of the response. In exploring these regimes we will embed the Mott result in a larger ``phase diagram'' in the $(\omega, E_0)$ plane.
 This phase diagram (Figure \ref{fig:regimesGD_lengthscales} (middle)) exhibits three new regions  which we characterize as exhibiting perturbative non-linear response, adiabatic non-linear response and enhanced dissipation. One central message of our analysis is that the even the limit of asymptotically small $\omega$ and $E_0$ in a localized system depends sensitively on the relative magnitudes of the two quantities.

At this point it is useful to distinguish our results from a more standard understanding of the limits of linear response in a more conventional solid state setting. In the latter setting one finds the same linear response, but the long time and large amplitude response will involve coupling to delocalized phonons in an essential manner. By contrast our results are intrinsic to the electronic system and 
probe the physics of Anderson localization alone, even outside the linear response regime.

In the main text we organize our discussion as follows. In Section \ref{sec-overview} we offer an overview of our results and introduce three length scales which organize the physics of linear and non-linear response. Section \ref{analysis} is the technical heart of the paper wherein we analyze the $(\omega,E_0)$ phase diagram using a combination of perturbation theory, Rabi oscillation theory, Landau-Zener tunneling ideas and Floquet theory. In Section \ref{sec-numerics} we present detailed numerical studies that bear out the ideas developed earlier. We close with a recapitulation of our main themes and results in Section \ref{sec-discuss}. In Appendix \ref{singlesite} and Appendix \ref{app-numerics} we discuss the case of a single site drive which has some useful pedagogical features.

Before presenting an overview of our results, we note that this work synthesizes and builds on many themes in recent work. Most narrowly it builds on the identification of the surprising, non-local, adiabatic response of localized insulators to a local perturbation in Ref.~\onlinecite{nonlocalrearrange}. More broadly it builds on work establishing the existence of Floquet many body localized systems\cite{Lazarides_2015,PonteHuveneers_2015, DimaHeating} which exhibit partially universal states at long times \cite{Khemani15, TCms}. In these many-body systems, the FLTS exhibits a reduction to the ``diagonal ensemble'' in which all observables vary periodically with the period of the drive and are said to synchronize with it\cite{FLTSnote}. For our non-interacting system global observables do synchronize, but local ones do not. When we study global energy absorption upon driving the system starting from a general state, there is a transient regime before a FLTS is reached. For weak driving and an initial equilibrium/ground state, this transient regime {\it is} the regime described by linear response theory. In our phase diagram, the transient response lasts less than one period of the drive except in the linear response regime; in all other regimes we will be discussing properties of the FLTS.

We note that an early version of these results was presented by one of the present authors~\cite{KITPtalks} and has also appeared in the DPhil thesis of another \cite{DTLthesis}. In the course of completing this work, there have been a few separate papers discussing response and regimes of energy absorption in driven MBL systems\cite{sarang, rehn, Kozarewski, DimaHeating}. In particular, our analysis shares many qualitative features with the discussion in Ref.~\onlinecite{sarang}, although we consider heating starting from a low-temperature initial state of the undriven Hamiltonian while Ref.~\onlinecite{sarang} works near infinite temperature. 
Despite qualitative similarities with the MBL case, the noninteracting problem offers a high degree of tractability which allows us to propose analytically and verify numerically several different independent signatures of the four different dynamical regimes suggested by our phenomenological analysis. While this is of interest in its own right, it also helps bolster the analogous analyses in MBL systems where numerics are limited to much smaller sizes and times. A key difference between the two cases is that the long time limit in interacting systems exhibits heating to infinite temperature for frequencies below a threshold set by local energetics\cite{Lazarides_2015,PonteHuveneers_2015, DimaHeating}. By contrast, in our disordered non-interacting system in one dimension, the energy absorption {\it always} saturates below the maximum possible value and the system enters a non-thermal late time state.

We also note related studies of non-interacting driven systems. In particular, we flag studies of tight-binding models which examine the effect of periodically driving a disordered one-dimensional system on localization length\cite{molina1,hatami,Ducatez2017}, conductance\cite{molina2, kitagawa}, spectral statistics\cite{molina3}, and localization properties of Floquet operator eigenstates \cite{bhatt17}. Ref~\onlinecite{bhatt17} also studied dissipative non-linear charge response (but not heating) of Anderson insulators, but they worked primarily in the strong drive limit $E_0\gg \omega$. 
Studies using random matrices as models for disordered physical systems have also asked questions related to our themes. Work on these models demonstrates the saturation of energy absorption in the presence of a drive when the Floquet eigenstates are localized\cite{wilkinson2} and examines the relationship between energy absorption and drive amplitude and frequency, depending on the class of random matrix\cite{wilkinson1, machida}. Additionally, Ref.~\onlinecite{cohen} studies the break down of linear response in a random matrix Hamiltonian under periodic driving. Separately, within linear response, the Mott law has been reproduced in numerical studies of a disordered one-dimensional tight-binding model \cite{saso}.

\section{Overview}\label{sec-overview}

In this section we outline a physical picture for the different regimes of behavior that arise in a periodically driven Anderson insulator, as the drive amplitude and frequency are varied. We focus on one-dimensional systems and consider frequencies low enough that the energy of a drive photon is much less than the spacing between electron energy levels in a region of size equal to the localization length $\xi$. The regimes of behavior are set partly by the relative sizes of three key length scales, which we introduce in the following and denote (in units of $\xi$) by $r_{\rm Mott}$, $r_L$ and $r_c$.

Linear response can by definition be described in terms of transitions between eigenstates of the undriven Hamiltonian, and the dominant contribution in the Anderson insulator at low frequency is from Mott resonances -- hybridized pairs of localized states with an energy splitting that matches the drive photon energy\cite{mott}. These pairs have a frequency-dependent characteristic spatial separation known as the Mott length, $r_{\rm Mott}$. This is the first of our three length scales. Within linear response theory, energy is absorbed by the sample from the drive at a constant rate. At finite but weak drive amplitude, response saturates on a timescale much longer than the drive period. The saturation can be understood by examining Rabi oscillations of the Mott resonances. The timescale to reach saturation decreases with increasing drive amplitude and a boundary to the regime of linear initial response is set by the amplitude at which the saturation time matches the drive period.

Other physical processes make contributions to the response that compete with Mott resonances as drive amplitude is increased. We discuss these processes by considering the eigenstates of the instantaneous Hamiltonian and their parametric variation over the drive period, taking the driving electric field to be represented using a scalar potential. A sufficient (but unnecessarily restrictive) condition for linear response theory to be valid at short times is that the variation of instantaneous energy levels over a period is much less than the level spacing. This is the case in a small system at weak drive. For a large system, however, there exist pairs of levels having energies that are close in the undriven system and are swept past each other by a finite amplitude drive. The drive amplitude determines a minimum spatial separation $r_L$ for the localization centers of such levels, which is the second of our key length scales. It is natural to consider these crossings using Landau-Zener theory. The quantum evolution during such a crossing depends on the strength of coupling between the levels and on the drive frequency, and can be characterized by our third length scale, $r_c$. Crossings between localized states with spatial separation much larger than $r_c$ are deep in the diabatic limit, while crossings between states with separation much smaller than $r_c$ are deep in the adiabatic limit. 

This distinction determines the contribution to the response of an Anderson insulator arising from the level crossing of a state occupied by an electron with another one that is initially empty. A strictly diabatic level crossing makes vanishing contribution because the electron does not move in space. Conversely, a strictly adiabatic crossing makes a contribution that is large and reactive, since the electron jumps between the localization centers of the two states involved, but jumps back later in the drive cycle when the two levels cross in the opposite sense. A dissipative response arises just from those crossings that are intermediate between diabatic and adiabatic, involving localized states with spatial separation of order $r_c$\cite{nonlocalrearrange, sarang, DimaHeating}. 

We provide estimates of these three length scales and their dependence on drive amplitude ${\cal E}$ in Sec.~\ref{analysis}. A schematic view of the results is given in Fig.~\ref{fig:regimesGD_lengthscales}. At weak drive the inequality $r_L \gg r_c \gg r_{\rm Mott}$ holds. In this first case, the only transitions that lie outside linear response theory (those between localized states with separation greater than $r_L$) are strictly diabatic and so unimportant. 
Above a critical drive strength the inequality is reversed, so that $r_{\rm Mott}\gg r_c \gg r_L$. In this second case Mott resonances are unimportant for response, because under drive they are traversed diabatically. Instead there is a reactive contribution to response, from pairs of localized states with spatial separation in the range between $r_L$ and $r_c$, and a dissipative contribution, from pairs with separation of order $r_c$. 

As illustrated in Fig.~\ref{fig:regimesGD_lengthscales}, we identify from this discussion four regimes of response for a one-dimensional Anderson insulator driven by a low-frequency oscillating electric field. Smooth crossovers between these regimes are traversed successively with increasing drive strength $\cal E$ at any fixed frequency $\Omega$. At the weakest drive strengths (the \emph{linear response} regime), linear response of Mott resonances dominates until a saturation time that is much longer that the drive period, and response after the saturation time is from Rabi oscillations of Mott resonances. The saturation time decreases with increasing drive strength, reaching the drive period at the upper boundary of the linear response regime. At higher drive strengths (the \emph{perturbative non-linear response} regime) there is no distinct period of initial response, and multi-photon transitions make a significant contribution to energy absorption. In both the linear response and perturbative non-linear response regimes, the ordering of length-scales is $r_L \gg r_c \gg r_{\rm Mott}$. This is reversed on entering the \emph{adiabatic non-linear response} regime, above a second threshold for drive strength: here there is a large reactive response from pairs of localized states with separation $r$ in the range $r_L \ll r \ll r_c$. In addition, at the highest drive strengths (the \emph{enhanced dissipation} regime) there is a large dissipative response from pairs with $r\sim r_c$.

\section{Detailed analysis}\label{analysis}

We study a one-dimensional or quasi one-dimensional Anderson insulator with density of states $\rho$ per unit length and energy, driven by an oscillating electric field of strength $E_0$ and frequency $\omega$. Let $e$ denote the electron charge. We use the inverse level spacing $ \xi\rho $ in a system of size $\xi$ to define the dimensionless field strength
\begin{equation}
{\cal E} \equiv eE_0\xi^2\rho
\end{equation}
and the dimensionless frequency
\begin{equation}
\Omega \equiv \hbar \omega \xi \rho\,.
\end{equation}
We are concerned with the response to weak fields (${\cal E} \ll 1$) at low frequencies ($\Omega \ll 1$) for a zero-temperature initial state in which the energy band of localized states in partially filled. We start by discussing the dependence on $\cal E$ and $\Omega$ of the characteristic lengths $r_{\rm Mott}$, $r_L$ and $r_c$ introduced above. 

Consider single-particle eigenstates in an Anderson insulator. Following Mott's picture of frequency-dependent conductivity, most eigenstates have a well-defined localization center, but a few form resonant pairs with other distant localized states. Energy absorption is due to transition within these resonant pairs. The minimum energy difference between the two states in a pair depends on their spatial separation $x$, because it is limited by level repulsion and this is controlled by spatial overlap between tails of wavefunctions. An estimate of this minimum energy difference is $(\rho \xi)^{-1} \exp(-x/\xi)$, and the Mott length is obtained by equating it to the energy $\hbar \omega$ of a drive photon. Introducing dimensionless lengths $r\equiv x/\xi$, the Mott length is therefore
\begin{equation}\label{neville}
r_{\rm Mott} = \ln(1/\Omega)\,.
\end{equation}

Pairs of states may be driven through a resonance by an electric field of finite strength. Representing the electric field using a scalar potential, the field modulates the relative energies of two states by an amount proportional to their spatial separation. The dimensionless length $r_L$ is defined by equating this energy modulation to the minimum energy separation of the pair. From $x e E_0 = (\rho \xi)^{-1} \exp(-x/\xi)$ we obtain at leading order for small $\cal E$
\begin{equation}
r_L \approx \ln(1/{\cal E})\,.
\end{equation}

Next we examine the dynamics of such an avoided crossing induced by an oscillating electric field. In general, time evolution of a pair of states $|m\rangle$ and $|n\rangle$ with the Hamiltonian ${\cal H}(t)$ and instantaneous energies $\varepsilon_n$ and $\varepsilon_m$ is adiabatic if 
\begin{equation}\label{adiabatic}
\frac{\hbar |\langle m| \partial_t {\cal H}(t) |n\rangle|}{(\varepsilon_m - \varepsilon_n)^2} \ll 1\,.
\end{equation}
For states with separation $x$ we set $|\langle m| \partial_t {\cal H}(t) |n\rangle| \sim eE_0x \omega$ and take $|\varepsilon_m - \varepsilon_n| \sim (\rho \xi)^{-1} \exp(-x/\xi)$. 
In this way the boundary between adiabatic and diabatic avoided crossings is located to be at ${\cal E} \Omega = r_c^{-1} \exp(-2r_c)$. For $\cal E$, $\Omega \ll 1$, this yields to leading order
\begin{equation}\label{rc}
r_c \approx \frac{1}{2} \ln\left(\frac{1}{{\cal E}\Omega}\right)\,.
\end{equation}
As illustrated in Fig.~\ref{fig:regimesGD_lengthscales}, at ${\cal E} = \Omega$ the lengths satisfy $r_{\rm Mott} \approx r_L \approx r_c$. For ${\cal E} \ll \Omega$ they have the ordering $r_L > r_c > r_{\rm Mott}$ and the effects of the electric field are perturbative.  For ${\cal E} \gg \Omega$ both inequalities are reversed, and some effects of the electric field are non-perturbative.

\subsection{Linear response}

In outline, a derivation of Mott's result for the frequency dependent conductivity in an Anderson insulator is as follows. We equate the macroscopic expression for the rate of energy absorption per unit length, in terms of the conductivity $\sigma(\omega)$, to a microscopic expression in terms of transitions between initial and final states $|{ i}\rangle$ and $|{f}\rangle$, with energies $\varepsilon_{ i}$ and $\varepsilon_{ f} = \varepsilon_{ i} + \hbar \omega$. Using the Fermi golden rule and denoting the density of final states by $\rho_{\rm F}$, this gives
\begin{eqnarray}
\frac{1}{2} E_0^2 \sigma(\omega) &=& \hbar \omega  \sum_{i} \rho \nu(\varepsilon_{i}) [1-\nu(\varepsilon_{ f})] \frac{2\pi}{\hbar} |\langle { i}|eE_0 x |{ f}\rangle|^2\rho_{\rm F} \,,\nonumber
\end{eqnarray}
where $\nu(\varepsilon)$ is the occupation probability of a state at energy $\varepsilon$. The central assumption is that the matrix element appearing here is small unless the initial and final states form one of the resonant pairs discussed above, in which case $|\langle { i}|eE_0 x |{ f}\rangle| \sim eE_0x_{\rm Mott}$. For the energy splitting of a pair to match the photon energy, the spatial separation of the pair should be within  ${\cal O}(\xi)$ of $x_{\rm Mott}$, and so in $d$ dimensions $\rho_{\rm F} \sim \rho x_{\rm Mott}^{d-1} \xi$. In consequence at zero temperature
\begin{equation}
\sigma(\omega) \sim \hbar e^2 \rho^2 \xi \omega^2 x_{\rm Mott}^{d+1}.
\end{equation}
In combination with the logarithmic dependence of $x_{Mott}$ on $\omega$ [Eq.~(\ref{neville})], this yields the expected result: Eq.~(\ref{MottLawEqn}).

\subsection{Rabi oscillations}\label{secRabi}

It is straightforward to treat long-time saturation of response at weak drive in terms of Rabi oscillations of resonant pairs. An effective Hamiltonian for one pair has the form
\begin{equation}
{\cal H}(t) = \left(\begin{array}{cc} \varepsilon_f & 0 \\ 0& \varepsilon_i \end{array}\right) + \left(\begin{array}{cc} 0&\gamma \\ \gamma & 0 \end{array}\right)\sin\omega t\,,
\end{equation}
where the coupling is $\gamma \sim eE_0 x_{\rm Mott}$. At finite drive strength one should take account of levels that are not exactly resonant. Denoting the detuning by $\delta = \varepsilon_f - \varepsilon_i - \omega$, the Rabi frequency is $\omega_{\rm R} = \sqrt{\delta^2 + \gamma^2}$ and the transition probability is 
\begin{equation}
P_{i\to f} = \frac{\gamma^2}{\delta^2+\gamma^2} \sin^2 (\omega_{\rm R}t/2) \,.
\end{equation}
We compute energy absorbed by summing $\hbar \omega P_{i\to f}$ over initial states $i$ with a spatial density $\rho \hbar \omega$, and averaging over $\varepsilon_f$ with energy density $\rho_{\rm F}$. For a sample of length $L$ this calculation gives the energy $\Delta E(t)$ absorbed at time $t$ as
\begin{equation}\label{linear}
\Delta E(t) \sim \left\{ \begin{array}{lcl} \tfrac{1}{2}L E_0^2 \sigma(\omega) t \sim \frac{L}{\rho \xi^2} {\cal E}^2 \Omega \ln^2(1/\Omega)\omega t & \, & t < t_{\rm sat}\\
&&\\
\tfrac{1}{2}L E_0^2 \sigma(\omega) t _{\rm sat} \sim \frac{L}{\rho \xi^2} {\cal E} \Omega^2 \ln(1/\Omega)& \, & t > t_{\rm sat}\,
\end{array}\right.
\end{equation}
where $\sigma(\omega)$ is in accord with Eq.~(\ref{neville}) and $t_{\rm sat} \sim \hbar/\gamma$.

\begin{figure}
    \begin{center}
	\includegraphics[width=0.47\textwidth]{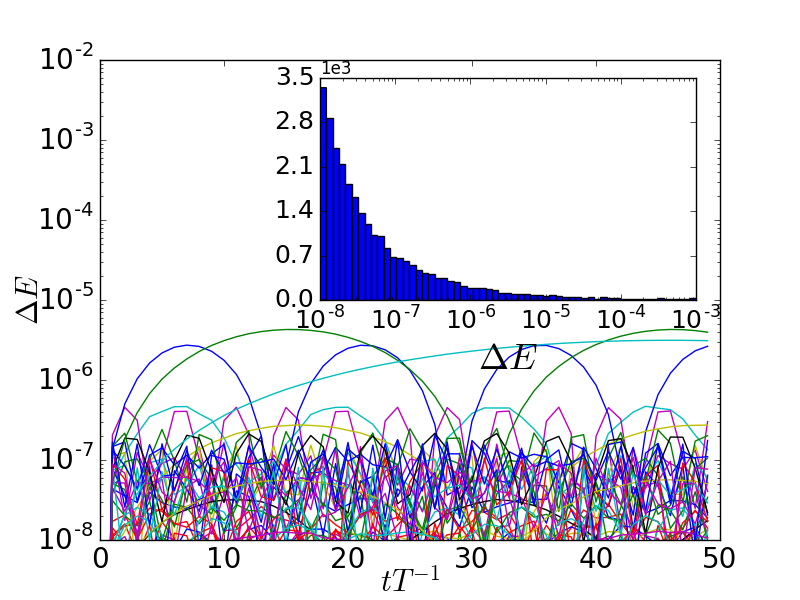}
	\caption{$\Delta E(t)$ vs $t$ for multiple disorder realizations. (inset) Histogram over disorder realizations of energy absorbed in long-time limit. Electric field strength, frequency, disorder strength, system size, and number of disorder realizations are, respectively: $\phi_0\omega = 2.5\times10^{-4}$, $\omega = 0.25$, $W=5$, $L=100$, and $N_r = 7.5\times 10^4$.}
	\label{fig:RealizationHistGD}
    \end{center}
\end{figure}

Rabi oscillations of isolated Mott resonances are vividly illustrated in numerical calculations of $\Delta E(t)$ vs $t$ for multiple disorder realizations in moderately sized samples, as shown in Fig.~\ref{fig:RealizationHistGD} [see Sec.~\ref{sec-numerics} for details of model and methods].  $\Delta E(t)$ in an individual realization exhibits Rabi oscillations. These vary widely in amplitude and period between realizations, with much larger energy absorption and longer period in resonant realizations than in typical ones.

\subsection{Perturbative non-linear response}

The boundary to the linear response regime is at $\omega t_{\rm sat} \sim 1$, which can be re-expressed in terms of the dimensionless field strength and frequency as 
${\cal E} \sim {\Omega}/{\ln(1/\Omega)}$.
In the limit $\Omega \ll 1$ of interest, there is a wide interval 
\begin{equation}
{\Omega}/{\ln(1/\Omega)}\ll {\cal E} \ll \Omega
\end{equation}
between the electric field strength at the boundary to the linear response regime, and the field strength at which the characteristic length scales $r_{\rm Mott}$, $r_L$ and $r_c$ cross. In this interval, the ordering of these lengths is $r_{\rm Mott}\ll r_c \ll r_L$. As a consequence, level crossings induced by the drive are strictly diabatic, and the leading correction to linear response theory arises from multi-photon absorption that can be treated using time-dependent perturbation theory at the appropriate order. This is the perturbative non-linear regime shown in Fig.~\ref{fig:regimesGD_lengthscales}.

\subsection{Strong field response}\label{sec:pertnlr}

At higher field strengths (${\cal E} \gg \Omega$) the system enters a new regime, in which avoided level crossings of the instantaneous Hamiltonian play a significant part in response. We argue in the following that time evolution in this regime can be understood in terms of a sequence of distinct two-level crossings.

As a first step, consider the avoided crossings involving a given level over a single period of the drive. We will estimate the average number $N_{\rm int}$ that are intermediate between diabatic and adiabatic, and the average number $N_{\rm ad}$ that are adiabatic. From the discussion leading to Eq.~(\ref{rc}), intermediate crossings occur between levels that have a spatial separation of their localization centers given by $\xi r_c$ to an accuracy ${\cal O}(\xi)$. The relative energies of two regions of the system with spatial separation $\xi r_c$ are modulated by $eE_0\xi r_c$ over a drive cycle. We therefore estimate
\begin{equation}
N_{\rm int} \sim eE_0 \xi^2 \rho r_c = {\cal E} r_c \sim {\cal E} \ln(1/{\cal E}\Omega)\,.
\end{equation}
Adiabatic avoided crossings occur between states with a spatial separation that lies between $\xi r_L$ and $\xi r_c$. Hence by a similar argument
\begin{equation}
N_{\rm ad} \sim \int_{\xi r_L}^{\xi r_c} eE_0 x \rho \, {\rm d}x \approx {\cal E} r_c^2 \sim {\cal E} [\ln(1/{\cal E}\Omega)]^2\,.
\end{equation}

To understand whether it is sufficient to consider only pairwise avoided level crossings, we estimate the fractions $F_{\rm int}$ and $F_{\rm ad}$ of the drive cycle occupied for a given level, by intermediate and adiabatic avoided crossings respectively. Taking the minimum energy spacing at an avoided crossing of levels with spatial separation $x$ to be $(\rho \xi)^{-1} \exp(-x/\xi)$, noting that the modulation in relative energy of two levels over a drive cycle is $eE_0 x$, and setting $x=\xi r_c$, we have
\begin{equation}
F_{\rm int} \sim N_{\rm int} \frac{(\rho \xi)^{-1}\exp(-x/\xi)}{eE_0 \xi r_c} \approx e^{-r_c}\,.
\end{equation}
A similar calculation gives
\begin{equation}
F_{\rm ad} \sim \int_{\xi r_L}^{\xi r_c}  \xi^{-1} \exp(-x/\xi) {\rm d} x \approx {\cal E}\,.
\end{equation}
Since $F_{\rm int} + F_{\rm ad}  \ll 1$ for ${\cal E}, \Omega \ll 1$, the avoided crossings involving a given level are mostly well separated from each other.

The regime of adiabatic non-linear response indicated in Fig.~\ref{fig:regimesGD_lengthscales} is the one in which $N_{\rm ad}>0$ but $N_{\rm int} \ll 1$, implying $\Omega \ll {\cal E} \ll [\ln(1/\Omega]^{-1}$, while the regime of enhanced dissipation is one with $N_{\rm int} \gg 1$, which requires $[\ln(1/\Omega)]^{-1} \ll {\cal E}$. We show below that in these two regimes there are characteristic contributions to energy absorption from intermediate crossings, and to the reactive response from adiabatic crossings. Rather surprisingly, we find that the dependence of these contributions on ${\cal E}$ and $\Omega$ does not change on crossing the boundary at $N_{\rm int}\sim 1$ between the two regimes.

\subsubsection{Adiabatic non-linear response}\label{LZ}

In the interval
\begin{equation}\label{bdy}
\Omega \ll {\cal E} \ll [\ln(1/\Omega]^{-1}
\end{equation}
only rare levels undergo intermediate avoided crossings.
In this regime energy absorption does not arise from Mott resonances, because they give rise to avoided crossing that in this range of drive strengths are traversed diabatically. Their place is taken by intermediate avoided crossings between levels with spatial separation $\xi r_L \ll x_{\rm Mott}$. Such an avoided crossing gives rise to energy absorption of order $eE_0 \xi r_L$ if it lies within $eE_0 \xi r_L$ of the Fermi energy. The spatial density of levels inside this energy window is $N_{\rm int} eE_0 \xi r_L\rho $, and so the total energy absorbed at long times is
\begin{equation}\label{sat}
\Delta E(\infty) \approx L N_{\rm int} (eE_0 \xi r_L)^2 \rho \approx (L/\rho\xi^2) {\cal E}^3 \ln^3(1/\Omega)\,.
\end{equation}
Note that this matches Eq.~(\ref{linear}) if it is evaluated at the boundary to the linear response regime. 

\subsubsection{Enhanced Dissipation}

The strongest range of drive strengths we consider is 
\begin{equation}
[\ln(1/\Omega)]^{-1} \ll {\cal E}.
\end{equation}
In this regime $N_{\rm int} \gg 1$, so that levels of the instantaneous Hamiltonian typically have many avoided crossings within a drive cycle that are intermediate between diabatic and adiabatic. Since $F_{\rm int} \ll 1$, these crossings occupy a small fraction of the drive cycle, and so it is appropriate to consider them in a pairwise fashion. 
We estimate the energy gain over a cycle by assuming that an electron in an eigenstate of the instantaneous Hamiltonian does a random walk in energy, consisting of $N_\text{int}$ steps, each of characteristic size $E_0\xi r_c$. Occupation is then spread over an energy window $\delta E$ and gives energy absorption $\Delta E \sim \rho L \lf \delta E \ri^2$. We expect $\delta E \sim eE_0 \xi r_c \sqrt{N_\text{int}}$, leading to the conclusion that Eq.~(\ref{sat}) applies in this region as well.

\subsection{Reactive response}\label{sec:reactive}

At all field strengths there is a reactive component to response, with contributions to $\Delta E(t)$ that oscillate over the drive cycle. The non-linear aspects of the reactive response are particularly interesting as they reflect the adiabatic transitions discussed above. A convenient way to compute the reactive component of $\Delta E(t)$, which we denote $\Delta E\lf t\ri_\text{reac}$, is via the polarization $P(t)$ of the sample in the presence of an electric field $E_0\sin \omega t$, since the current flowing is $I(t)=\partial P(t)$, which in turn is related to $\Delta E\lf t\ri_\text{reac}$ by $\partial_t \Delta E\lf t\ri_\text{reac} = I(t) E_0 \sin\omega t$.

\subsubsection{
Linear response}

In linear response, one has $P(t) = \chi E_0\sin \omega t$, where $\chi$ is the polarizability, and so
\begin{equation}
\Delta E\lf t\ri_\text{reac} = \frac{1}{2} \chi E_0^2 \sin^2 \omega t\,.
\end{equation}
The low-frequency polarizability is well approximated by its static value, given in terms of single particle eigenstates $|m\rangle$ and energies $\varepsilon_m$ for a system with chemical potential $\mu$ by
\begin{equation}
\chi = e^2 \sum_{ \varepsilon_m<\mu< \varepsilon_n} \frac{|\langle n|x|m\rangle|^2}{\varepsilon_n - \varepsilon_m}\,.
\end{equation}
We estimate the value of this expression by taking $|\langle n|x|m\rangle|\sim \xi$ if the two states are localized in a region of size $\xi$ and lie within an energy window of width $(\xi\rho)^{-1}$, and zero otherwise. This gives $\chi \sim e^2 \rho \xi^2 L$ for a one-dimensional system of size $L$. (The relation between this result and Mott's law via the Kramers-Kronig relations is discussed in Ref.~\onlinecite{Feigelman}.) Hence 
\begin{equation}
\Delta E\lf t\ri_\text{reac} \sim (L/\rho\xi^2) {\cal E}^2 \sin^2 \omega t\,.
\end{equation}

\subsubsection{Adiabatic non-linear response}

In the field range 
\begin{equation}
\Omega \ll {\cal E} \ll [\ln(1/\Omega)]^{-2}
\end{equation}
$N_{\rm ad} \ll 1$ so that adiabatic levels crossings are rare and an individual level is involved in at most a single crossing. We estimate $I(t)$ under these conditions as follows. Electrons which make adiabatic transitions at unit rate between states separated by distance $x$ contribute current $ex$. For an electron at energy $\varepsilon$ below the Fermi energy to make a transition to an empty state, the possible spatial separation range, $x$, is $x\geq x_{\rm min} \equiv \varepsilon/\lf eE_0 \sin \omega t\ri$ to ensure the final state is unoccupied, and $x\leq \xi r_c$ for the transition to be adiabatic. The rate at which states at distance $x$ pass through avoided crossings with a given initial state is $\rho e E_0 \lvert x \rvert \omega \cos \omega t$. Accounting for all possible initial states, for the first quarter of the drive cycle we have
\begin{align}\label{counting}
\partial_t \Delta E\lf t\ri_\text{reac} &= L\rho^2\lf eE_0\ri^2 \omega \sin \omega t\cos \omega t \int_0^{\varepsilon_{\rm max}} \!\!\!\!d\varepsilon \int_{x_{\rm min}}^{\xi r_c} \!\!\!\!x^2 dx\n\\
 &= \frac{\omega}{4} \lf r_c \xi e E_0\ri^3 L \rho^2 r_c\xi\sin^2 \lf \omega t\ri \cos\lf \omega t\ri.
\end{align}
where $\varepsilon_{\rm max} \equiv \xi r_c e E_0 \sin \omega t$. 

Accounting for the entire drive cycle and integrating over time, we find a reactive contribution to energy absorption at intermediate field strengths given by
\begin{align}\label{IntReacEqn}
\Delta E\lf t\ri_\text{reac} &\sim \lf \frac{L}{\rho\xi^2}\ri {\cal E}^3 \ln^4 \lf \frac{1}{\Omega}\ri\lvert\sin^3\lf \omega t\ri\rvert.
\end{align}

\subsubsection{Reactive response in the enhanced dissipative regime}

For the largest field strengths,
\begin{equation}
[\ln(1/\Omega)]^{-2} \ll {\cal E}\,,
\end{equation}
$N_{\rm ad} \gg 1$ and so individual levels are typically involved in many adiabatic crossings over a drive cycle. One might expect this to lead to a modification of Eq.~(\ref{IntReacEqn}), but surprisingly it does not, as we now show.

To capture the fact that there are many adiabatic crossings per level, we consider occupation $n\lf \varepsilon, t\ri$ of levels as a function of energy $\varepsilon$ and time $t$ within a cycle. The generalization of Eq.~(\ref{counting})  above is
\begin{equation}
\partial_t \Delta E\lf t\ri_\text{reac} = L\rho^2 \lf eE_0\ri^2 \omega \sin \omega t \cos \omega t\times  K
\end{equation}
with
\begin{align}
K&=\int_{-\infty}^{\infty} \!\!\!\!d\varepsilon \int_{-\xi r_c}^{\xi r_c} \!\!\!n\lf \varepsilon, t\ri \left[1 - n \lf\varepsilon + eE_0x\sin \omega t,t \ri  \right]\lvert x \rvert x dx \n\\
&=  \int_{-\infty}^{\infty} ds \int_{-1}^{1} \nu \lf s, t\ri \left[1- \nu \lf s+\ell, t\ri\right]\lvert \ell \rvert \ell d\ell .
\end{align}
Here we have substituted $\varepsilon = s r_c \xi e E_0 \sin \omega t, x = r_c \xi \ell$ and $\nu \lf s,t\ri \equiv n \lf sr_c \xi e E_0 \sin \omega t ,t\ri$. We can recover Eq.~(\ref{IntReacEqn}) by taking $\nu \lf s,t\ri = \Theta \lf -s\ri$.

In general, $\nu\lf s, t\ri$ should be a monotonic function that varies from $\nu(s,t) = 1$ at large, negative $s$ to $\nu(s,t) = 0$ for large, positive $s$, with a step of width $w$.  The integral $K$ is ${\cal O}(1)$ independently of $w$, and so the reactive contribution to response is given by Eq.~(\ref{IntReacEqn}) throughout the range of field strengths $\Omega  \ll {\cal E}$. 

\subsection{Evolution Operator}\label{evo}

To illustrate directly the consequences of our Landau-Zener picture for time evolution, we turn to a discussion of the evolution operator in the basis of instantaneous eigenstates. Let $\left\{ \lvert\varphi_k\lf t\ri\rangle\right\}$ be eigenstates of ${\cal H}\lf t\ri$. We define
\begin{align}\label{OverlapMatEqn}
S_{jk} \lf t, n\ri = \langle \varphi_j\lf t \ri \lvert \mathrm{T}_t \exp \left[-i \int_{0}^{nT+t} {\cal H} \lf t'\ri dt'\right] \rvert \varphi_k\lf 0\ri\rangle,
\end{align}
where $\mathrm{T}_t$ denotes time ordering. 

For the discussion that follows, the relative ordering of the labels $j$ and $k$ for instantaneous eigenstates of ${\cal H}(t^\prime)$ at times $t^\prime=0$ and $t^\prime = nT+t$ is central. While we have chosen to present our discussion of time-evolution in terms of avoided level crossings for a system with an electric field represented using a scalar potential, the most revealing ordering of $j$ and $k$ is by the instantaneous eigenvalues of ${\cal H}(t)$ with electric field represented using a time-dependent vector potential, because this preserves the real-spacing ordering of $\langle \varphi_j(t)|x|\varphi_j(t)\rangle$ and $\langle \varphi_k(t)|x|\varphi_k(t)\rangle$ in a transition that is diabatic. For a sample without periodic boundary conditions, either gauge choice is of course permissible and equivalent to the other provided time evolution is computed exactly. However the standard condition [Eq.~(\ref{adiabatic})] for an avoided level crossing to be adiabatic \emph{is} gauge-dependent and needs to be used with a scalar potential which ensures that a time independent field is represented by a time independent Hamiltonian as assumed in the derivation.
Finally, although a vector potential must be used to represent an oscillating electric field in a sample with periodic boundary conditions, sensitivity to boundary conditions is small provided $L \gg \xi$. 

If all level crossings are perfectly diabatic, $S_{jk}(t,n)$ is a diagonal unitary matrix for all $t,n$. If the level crossings are a mixture of perfectly diabatic and perfectly adiabatic, $S_{jk}(t,n)$ is the product of a permutation matrix and a diagonal unitary matrix. Furthermore, in the latter case, the permutation returns to the identity at $t=T/2$. Deviations from this behavior arise from intermediate level crossings, which are rare in the adiabatic non-linear regime. We will use $S_{jk}(t,n)$ to identify the adiabatic non-linear regime in numerical simulations.

In particular, we introduce the quantities
\begin{equation}\label{OverlapEqns}
f\lf t\ri \equiv L^{-1} \sum_k \langle \lvert S_{kk} \lf t,n\ri\rvert^2\rangle
\end{equation}
and
\begin{equation}
g\lf t\ri \equiv L^{-1} \sum_k \langle \max_{j} \lvert S_{jk} \lf t,n\ri\rvert^2\rangle
\end{equation}
for a system of $L$ states, averaged over disorder and $n$. We expect $1-g\lf t \ri \sim {\cal O} \lf N_\text{int}\ri$ and $1-f\lf 0 \ri \approx 1-f\lf T/2\ri \sim {\cal O}\lf N_\text{int}\ri$. At $t\neq 0,T/2$, $1- f\lf t\ri$ is a measure of the fraction of levels that have undergone adiabatic crossings. If $N_\text{ad} \gg 1$, we expect $f\lf t\ri \ll 1$ unless $t$ is near an integer multiple of $T/2$.

\section{Numerical Simulations}\label{sec-numerics}

In this section we present results from numerical simulations of a site-disordered, one-dimensional tight-binding model for spinless fermions driven globally by an electric field.

We start from the Hamiltonian
\begin{align}
H_0 &= - \sum_i \lf \lambda c_i^{\dag}c_{i+1} + \text{h.c.}\ri + \sum_i w_i c_i^{\dag}c_i,
\end{align}
where the site potentials $w_i$ are independent random variables uniformly distributed in $\left[-W,W\right]$. Representing the electric field using a time-dependent vector potential introduced via the Peierls substitution, the model with global drive is
\begin{equation}
H_\text{GD}\lf t \ri   = -\sum_i \lf \tilde{\lambda}\lf t\ri c_i^{\dag}c_{i+1} + \text{h.c.}\ri +\sum_i w_i c_i^{\dag}c_i\,,
\end{equation}
where $\tilde{\lambda}\lf t\ri         = \lambda\exp\lf -i\phi_0 \cos\lf \omega t \ri\ri$. The product of charge and electric field strength is then $eE(t) = \phi_0 \omega \, \sin \omega t$. 

Taking $\lambda=1$ to set energy scales, we focus on strong disorder, with $W = 2$, $5$, $10$, and $20$. For $W=2$, $\xi \approx 6$ lattice spacings and for $W = 5,10$, and $20$, we have that $\xi \leq 1$ lattice spacing \cite{kappus}. We consider frequencies in the range $2.5 \times 10^{-4}\leq \omega \leq 0.5$ and drive strengths, $\phi_0\omega$, in the range from $10^{-5}$ to $1$. These correspond to dimensionless field strength and frequency in the ranges $10^{-6} \lesssim {\cal E} \lesssim 10^{-2}$ and $10^{-5} \lesssim \Omega \lesssim 10^{-1}$, respectively. We use as an initial state the ground state of a system with the Hamiltonian evaluated at $t=0$ and zero chemical potential. We consider systems with periodic boundary conditions and $L=100$ sites except where otherwise noted. Most results are averaged over $N_r$ disorder realizations, with 
 $10^4 \leq N_r \leq 5\times 10^5$.
 
To compute time-evolution numerically, we construct the time-evolution operator $U\lf t,0\ri$ from a piecewise-constant, discretized version of $H\lf t\ri$, using $N_\delta$ time steps in a period. Typically $N_\delta\sim 2\times 10^2$ is adequate, but for small frequencies values as large as $N_\delta = 2\times10^5$ are necessary; for further discussion, see Ref~\onlinecite{DTLthesis}.

In the following, we present results for range of quantities that characterize behavior. We consider energy absorption, changes in fermion occupation of eigenstates of the initial Hamiltonian and of lattice sites, and fluctuations of these quantities. Additionally, we study some of the observables discussed in Sec.~\ref{analysis}: we investigate the dependence on electric field strength of different harmonics of the reactive and dissipative contributions to energy absorption within the drive cycle; and we compute the overlap matrix within the drive cycle, as defined in Eq.~(\ref{OverlapMatEqn}). We focus on systems with global drive, but reference results for a local drive where the comparison is illuminating. Data for systems with a single-site drive are given in Appendix~\ref{app-numerics}.

\subsection{Energy absorption}
\begin{figure*}
    \includegraphics[width=0.33\textwidth, height = 0.3\textwidth]{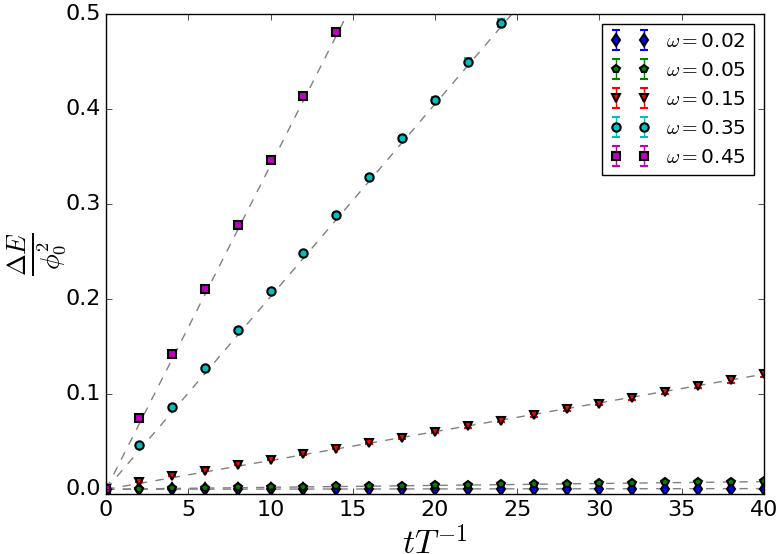}
    \label{fig:LRAbsGD}
    \raisebox{0.015\height}{\includegraphics[width=0.32\textwidth, height = 0.29\textwidth]{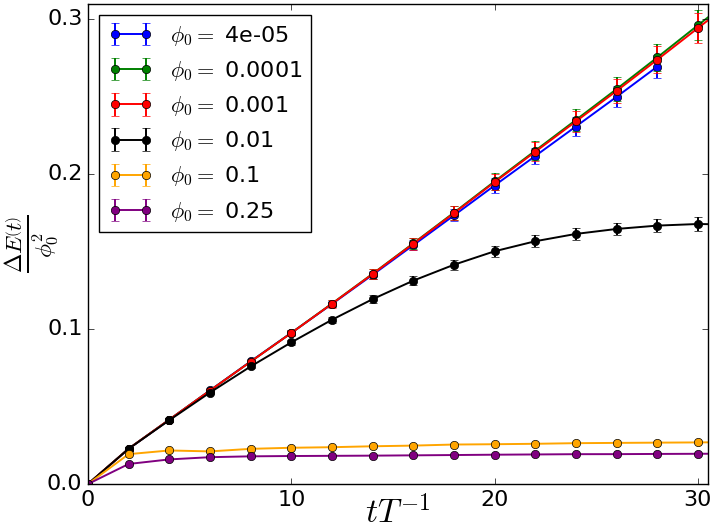}}
    \label{fig:AmpSquaredLinearAbsGD}
    \includegraphics[width=0.33\textwidth, height = 0.3\textwidth]{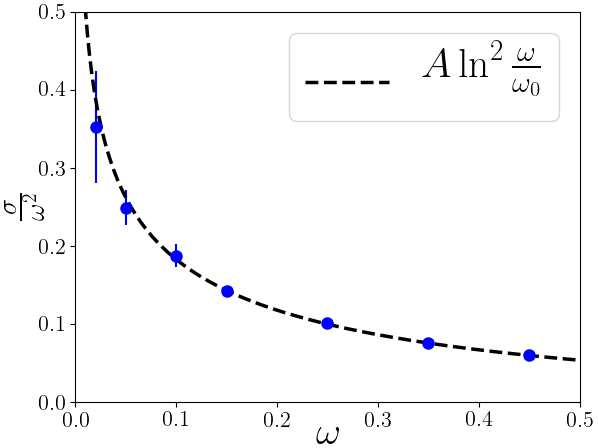}
    \caption{(left) Absorbed energy vs time at weak drive, showing $\Delta E(t) \propto t$ over many drive periods $T$. Parameter values: $\phi_0=10^{-4}$. Dashed lines are linear fits: points are data measured every two periods (left and middle). (middle) $\Delta E(t)/\phi_0^2$ vs $t$ for a range of drive strengths, demonstrating a short-time regime characterized by a conductivity that is independent of drive strength, and saturation at long times. Lines are guides to the eye. Parameter values: $\omega = 0.25$. (right) Comparison of frequency dependence of conductivity with Mott law: $\sigma/\omega^2$ vs $\omega$. Points: data analysed using Eq.~(\ref{sigma}); dashed curve: fit to Mott law, Eq.~(\ref{mott}), with $A \approx 0.016$, $\omega_0\approx 3.3$. (all) Other parameters values: $W=5$ and $N_r \sim 1 \-- 5\times 10^5$.}
    \label{fig:MottLawCondGD}
    \label{fig:TwoColGD}
\end{figure*}

An overall characterization of the response of a system to a periodic drive is given by the normalized energy absorption 
\begin{align}
\Delta E\lf t\ri &\equiv \frac{\langle\psi\lf t\ri \vert H\lf t \ri\vert\psi \lf t\ri\rangle - E_0}{E_\infty - E_0}
\end{align}
at integer multiples of the drive period. Here $E_\infty$ is the energy at infinite temperature and $E_0$ is the energy at $t=0$. 

We start with a demonstration of the expected behavior in the linear response regime. Specifically we show that: (a) $\Delta E(t)$ is initially proportional to $t$ over an interval that extends to times much larger than the period $T$ when the drive amplitude is weak; (b) the energy absorption rate in this interval is quadratic in the drive amplitude; (c) the frequency-dependence of this rate (proportional to the conductivity) is consistent with predictions of variable-range hopping; and (d) the timescale $t^*$ at which energy absorption saturates, and the energy $\Delta E(\infty)$ absorbed at long times, both have a dependence on drive amplitude and frequency consistent with predictions from our treatment of Rabi oscillations of driven Mott resonances.

Evidence is provided for (a) in Fig.~\ref{fig:TwoColGD} (left) and for (b) in Fig.~\ref{fig:TwoColGD} (middle). To examine (c) we extract from the weak-field, short-time behavior 
\begin{equation}\label{sigma}
\Delta E(t) = \sigma \,2\pi (\phi_0 \omega)^2  t
\end{equation}
the coefficient $\sigma$. As shown in Fig.~\ref{fig:TwoColGD} (right), its dependence on frequency matches well the Mott law expectation
\begin{equation}\label{mott}
\sigma = A \omega^2 \ln^2(\omega/\omega_0)\,.
\end{equation}
It is interesting to contrast this Mott law behavior for a global drive with the corresponding result for a single-site drive, where from App.~\ref{singlesite} one expects $\sigma = B \omega^2$ \emph{without} the characteristic $\ln^2(\omega/\omega_0)$ factor. This difference is apparent in a comparison of Fig.~\ref{fig:TwoColGD} with Fig.~\ref{fig:TwoColSSD}.

\begin{figure}
    \begin{center}
	\includegraphics[width=0.47\textwidth]{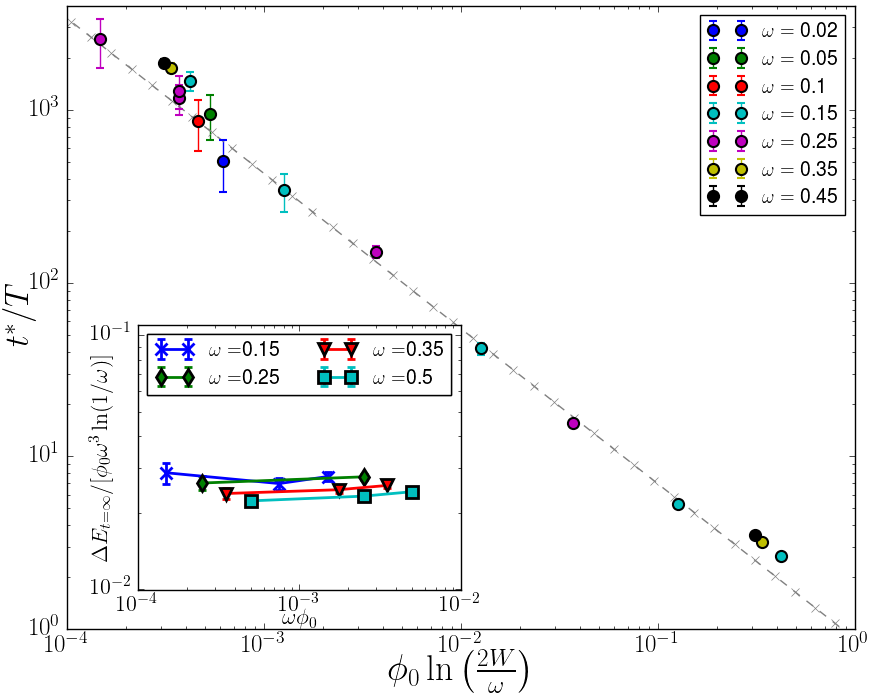}
	\caption{Dependence of saturation time on drive strength and frequency: $t^\star/T$ vs $\phi_0 \ln(2W/\omega)$ on log-log scales. Points: data; dashed line: fit with slope $-1$.  
	Parameter values: $W=5$ and  $N_r \sim 2 \times 10^5$. (inset) Dependence of $\Delta E(\infty)$ on $\omega$ and drive strength, $\omega \phi_0$.
	Ratios of $\Delta E(\infty)$ to predicted $\omega$-dependence [Eqn.~(\ref{Einfty})] vs $\omega$ on log-log scales. Points: data; lines: guides to the eye. 
	Parameter values: $W=5$, $N_r \sim 1 \times 10^4 \-- 1\times 10^5$.}
	\label{fig:TstarGD}
    \end{center}
\end{figure}

At long times, energy absorption saturates. It is clear from Fig.~\ref{fig:TwoColGD} (middle) that the timescale for saturation varies with drive strength. To quantify this dependence we define the saturation time $t^*$ to be the time at which $\Delta E(t)$, extrapolated linearly from its short-time behavior, 
reaches $\Delta E(\infty)$. From the theory of Rabi oscillations presented in Sec.~\ref{secRabi}, we expect
\begin{equation}\label{t^*}
\frac{t^*}{T} = \frac{1}{2\pi \phi_0 \xi \ln(2W/\omega \xi)}
\end{equation}
and
\begin{equation}\label{Einfty}
\Delta E(\infty) = \frac{L\xi^2}{4W^2} \phi_0 \omega^3 \ln(2W/\omega \xi)\,.
\end{equation}
Evidence in support of Eq.~(\ref{t^*}) is provided by the collapse of data to the dashed line in Fig.~\ref{fig:TstarGD}. Similarly, the data collapse shown in the inset of Fig.~\ref{fig:TstarGD} is consistent with Eq.~(\ref{Einfty}). 

It is again interesting to contrast these results for global drive with the different functional forms for $t^*$ and $\Delta E(\infty)$ in the case of single-site drive. Results from the theory of Rabi oscillations in that case are given in Eqns.~(\ref{t^*app}) and (\ref{Einftyapp}), while data is shown in Fig.~\ref{fig:TstarSSD}.

We note finally that the boundary of the linear response regime, as discussed in Sec.~\ref{sec-overview} and sketched in Figs.~\ref{fig:regimesGD_lengthscales} (middle) and \ref{fig:regimesSSD}, is set by $t^* \sim T$. Since simulations for both single-site and global drive give results for $t^*$ that are consistent with theoretical expectations, so is the location of this boundary.

\subsection{Energy distribution of excitations}
\begin{figure*}
    \includegraphics[width=0.32\textwidth, height = 0.29\textwidth]{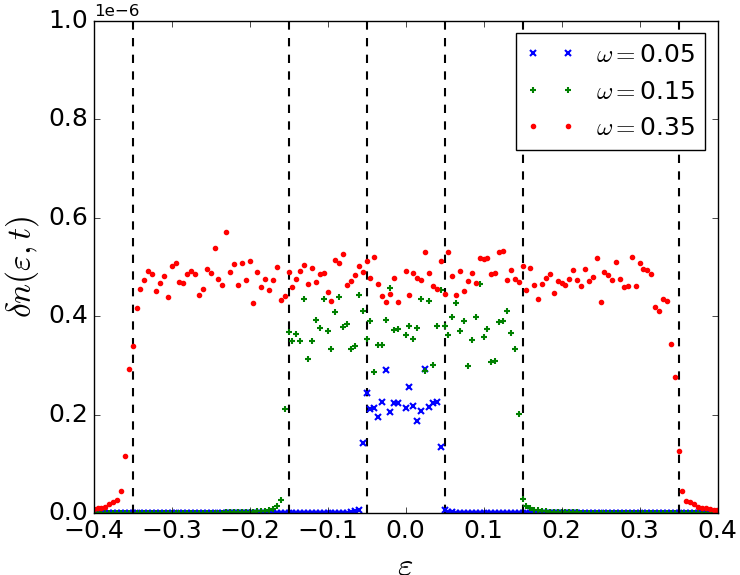}
    \label{fig:LRQuantaGD}
    \includegraphics[width=0.33\textwidth, height = 0.29\textwidth ]{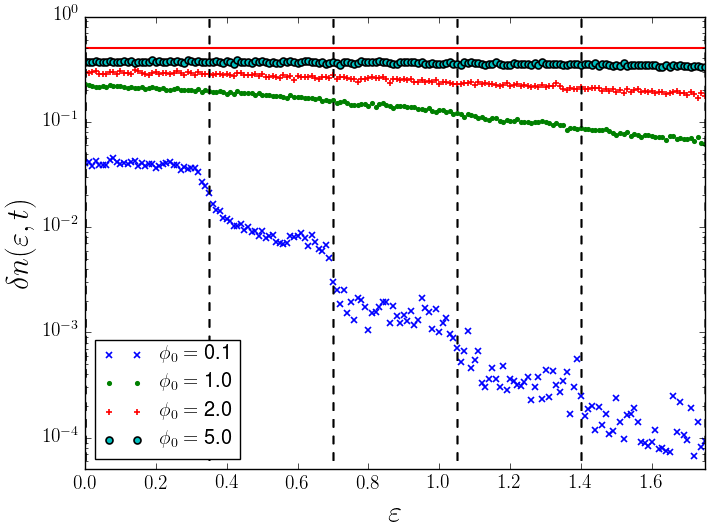}
    \label{fig:NLRQuantaGD}
    \includegraphics[width=0.32\textwidth, height = 0.31\textwidth]{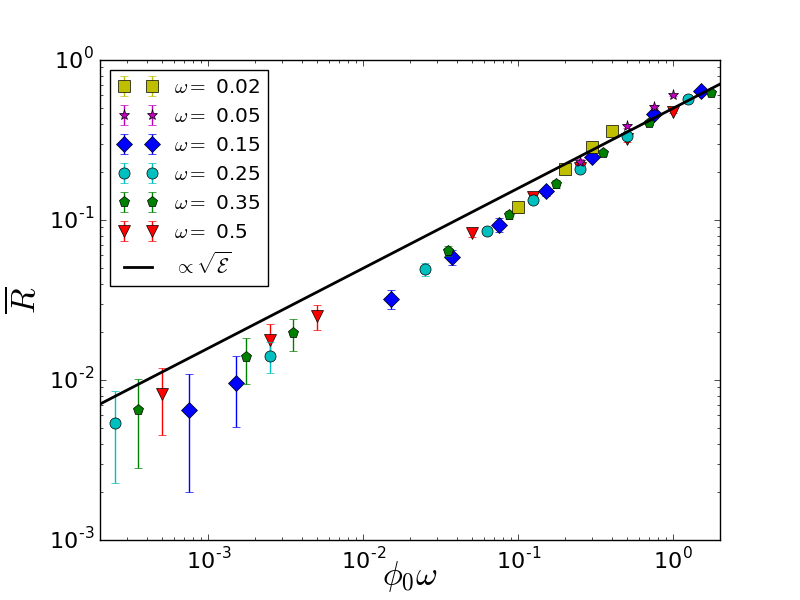}
    \caption{(left) Changes in eigenstate occupation induced by weak drive: $\delta n(\varepsilon,t)$ vs $\varepsilon$ at $t=20T$. Excitations are present only inside the energy window $|\varepsilon| < \omega$. Parameter values: $\phi_0 = 10^{-4}$. (middle) Changes in eigenstate occupation induced by intermediate and strong drive: $\delta n \lf \epsilon, t\ri$ vs $\varepsilon$ at $t=\infty$. Parameter values: $\omega = 0.35$. (right) Dependence of $\overline{R}$ [see Eq.~(\ref{Rbar})] on drive amplitude $\phi_0\omega$ with log-log scales. Points: data. Line: $\overline{R} \propto \sqrt{\phi_0 \omega}$, as expected at strong drive. (all) Other parameter values: $W = 5$, $N_r \sim 1\times 10^4 \-- 5\times 10^5$.}
    \label{fig:SampActGD_powerlaws}
    \label{fig:QuantaActGDCombined}
\end{figure*}
In order to expose the microscopic physics behind energy absorption, it is interesting to examine how the occupation of eigenstates of $H(0)$ changes during time evolution. Let $n(\varepsilon,t)$ be the fermion occupation number of an eigenstate with energy $\varepsilon$ at time $t$. From our choice of initial state, we have $n\lf \epsilon, 0\ri =\Theta\lf -\epsilon\ri$.
We write the change in occupation, relative to $t=0$, as
\begin{align}\label{OccupationEqn}
\delta n\lf \epsilon,t\ri &\equiv \begin{cases} 
                    1 - n\lf \epsilon,t\ri &\epsilon < 0\\
                    n\lf \epsilon, t\ri &\epsilon > 0\,.
                                  \end{cases}
\end{align}

In the linear response regime, energy absorption is expected to arise from resonant pairs of states. These states are separated by energy $\omega$. Due to Pauli exclusion, only occupied states within $\omega$ of the Fermi energy can be excited. We therefore expect a depletion in the occupation of states with $-\omega < \epsilon < 0$ and an excess occupation of states with $0 < \epsilon < \omega$, but no change in occupation for $|\varepsilon| > \omega$, so that $\delta n(\varepsilon)$ is positive for $|\epsilon|<\omega$ and zero otherwise.  Fig.~\ref{fig:QuantaActGDCombined} (left) displays exactly this behavior at weak drive. Fig.~\ref{fig:QuantaActGDCombined} (middle) illustrates behavior as drive strength is increased: at intermediate drive strength ($\phi_0 = 0.1$) multi-photon absorption is apparent, while at higher drive strengths ($\phi_0 \geq 1$) no structure is visible in the energy dependence of $\delta n \lf \epsilon, \infty\ri$.

\subsection{Spatial distribution of excitations}

A second way to illustrate the physics of Mott resonances at weak drive, and to investigate new features at strong drive, is to examine the spatial distribution of particle excitations. Let $n(x,t)$ be the expectation value of the fermion number operator at site $x$ and time $t$ in a given realization of the driven system, and let $\delta n(x) = n(x,\infty) - n(x,0)$. We compute
\begin{align}\label{Rbar}
\overline{R} &= \frac{\langle\lf\delta n(x)\ri^2 \rangle}{\langle \lf\delta n(x)\ri^4\rangle^{1/2}}\,
\end{align}
where the averages are over sites $x$ and disorder realizations.
This ratio characterizes the fraction of the system that has significant change in occupation. For example, if $\lf\delta n(x)\ri^2 = 1$ on a fraction $f$ of sites and is zero elsewhere, then $\overline{R}= f^{1/2}$. At weak drive, we expect only resonant pairs are active so that $\overline{R}$ is small. For strong drive, the system is more uniformly active and so $\overline{R}$ increases towards one. Fig.~\ref{fig:QuantaActGDCombined} (right) illustrates just this behavior. From the arguments of Sec.~\ref{sec:pertnlr} we expect at sufficiently strong drive $\overline{R} \propto \sqrt{N_{\rm ad}}\sim \sqrt{\cal E}$, and the data are consistent with this.

\subsection{Harmonics of reactive and dissipative response}

As discussed in Secs.~\ref{sec:pertnlr} and \ref{sec:reactive}, we expect different dependence on ${\cal E}$ of the reactive and dissipative contributions to response as field strength is varied. To examine these differences we consider a Fourier decomposition of the energy absorption at long times within a drive cycle
\begin{align}\label{DiscFourierEqn}
\Delta E \lf t \ri &= \sum_n c_n e^{in\omega t}.
\end{align}

\begin{figure*}
	\includegraphics[width=\textwidth]{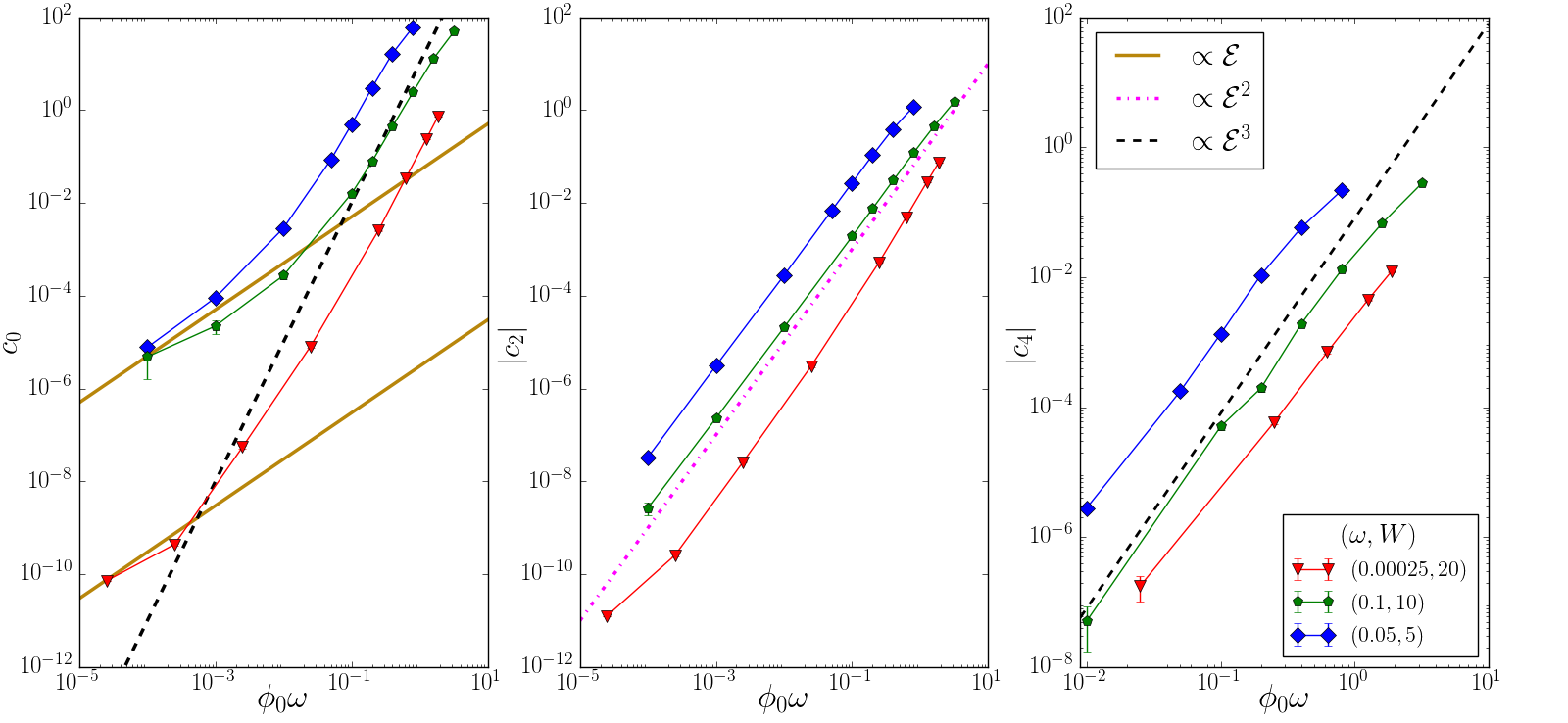}
	\caption{(Global drive) Fourier coefficients $ c_0, \lvert c_2 \rvert, \lvert c_4 \rvert$ of $\Delta E\lf t \ri$ vs drive strength on log-log scales 
	Points: data; lines: expected power laws (see main text).
	Parameter values: $L=26 \-- 48$, $N_r \sim 5\times 10^4 \-- 1.5\times 10^5$.}
	\label{fig:FourierModesGD}
\end{figure*}

The amplitude of the dissipative response is $c_0$. From Sec.~\ref{analysis} we expect $c_0 \propto {\cal E}$ at weak field and $c_0 \propto {\cal E}^3$ at strong field. Fig.~\ref{fig:FourierModesGD} (left) shows this crossover. It occurs at an electric field strength that increases with frequency; in addition, 
$c_0 $ is strongly dependent on frequency in the linear response regime, but much less so in the strong driving regime. These features are agreement with the predictions of Eqns.~\ref{linear}, \ref{bdy} and \ref{sat}. 

The values of $\lvert c_2 \rvert$ and $\lvert c_4 \rvert$ characterize the reactive response. The amplitude of polarization oscillations induced at the fundamental frequency by the drive field is given by $\lvert c_2 \rvert$. These oscillations arise in linear response and so one expects  $\lvert c_2 \rvert\propto {\cal E}^2$. This behavior is seen in Fig.~\ref{fig:FourierModesGD} (middle). Higher harmonics of the reactive response arise at strong field. From Landau-Zener theory of the evolution operator, we expect
$\lvert c_4 \rvert \propto {\cal E}^3$, as found in Fig.~\ref{fig:FourierModesGD} (right).~\cite{NominalContributionNote} Note that with our choice of phase for the driving field, the odd Fourier components of $\Delta E(t)$ are zero.

\subsection{Evolution operator at strong drive}
\begin{figure}
    \begin{center}
	\includegraphics[width=0.47\textwidth]{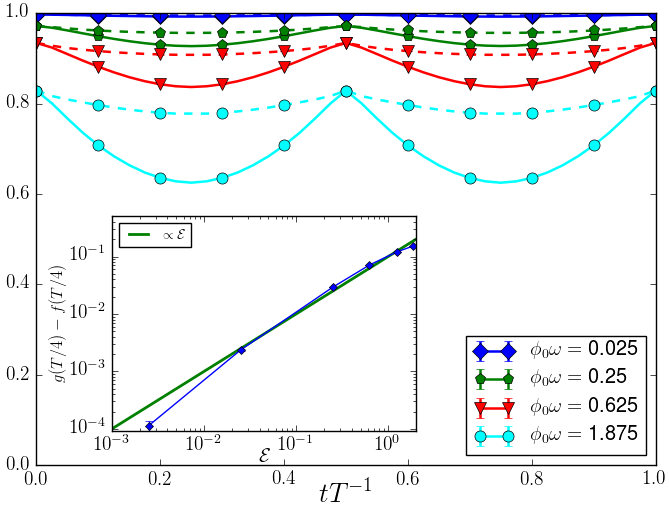}
	\caption{Variation with $t$ of functions $f\lf t\ri$ (solid) and $g\lf t\ri$ (dashed)  over one drive period at long times [see Eq.~(\ref{OverlapEqns}) for definitions]. Adiabatic level crossings lead to a reduction in the value of $f\lf t\ri$ when $t\neq T/2$. Values of $g\lf t\ri$ less than $1$ arise from level crossings that are intermediate between adiabatic and diabatic. Parameter values: $\omega = 2.5\times 10^{-4}$, $L=26$, $W=20$, $N_r \sim 1.5\times 10^5$. (inset) Variation of $g\lf T/4\ri - f\lf T/4\ri$  with ${\cal E}$ on log-log scale in the strong drive regime. Points: data; line: linear dependence from Landau-Zener theory omitting log corrections. Parameter values: $\omega = 2.5\times 10^{-4}$, $L=26$, $W=20$, $N_r\sim 1.5\times 10^5$.}
	\label{fig:OverlapsGD}
    \end{center}
\end{figure}

Next we present the results of calculations designed to test the theory of the evolution operator at strong drive, developed in Sec.~\ref{LZ} using a picture of Landau Zener crossings of pairs of localized eigenstates of the instantaneous Hamiltonian.
Specifically, we study the quantities $f\lf t \ri$ and $g\lf t\ri$, introduced in Eq.~(\ref{OverlapEqns}), over a Floquet period at long times. 
In the adiabatic non-linear regime, 
we expect that $f\lf t\ri \ll 1$ except when $t$ is near an integer multiple of $T/2$ and that $g\lf t \ri \sim 1 - {\cal O} \lf N_\text{int}\ri \approx 1$ throughout the drive period.
Numerical constraints make the regime in which $N_\text{ad} \gg 1$ difficult to access. 
Nonetheless, in Fig.~\ref{fig:OverlapsGD} the expected behavior is apparent. 

An additional test of the application of Landau-Zener theory to the evolution operator is provided by the dependence of the amplitude of the oscillations in $f(t)$ on drive strength. The difference $g\lf T/4\ri - f\lf T/4\ri$ is expected to be proportional to $N_\text{ad}-N_\text{int}$ and vary as ${\cal E}$ for strong drive. Evidence for this behavior is provided in the inset of Fig.~\ref{fig:OverlapsGD}.

Note that the data shown in Figs.~\ref{fig:FourierModesGD}, \ref{fig:OverlapsGD} is intended to probe behavior deep in the adiabatic non-linear regime. Access to this regime required higher disorder strength, lower frequency, and finer discretization of the time-evolution operator ($N_\delta \sim 10^5$) than data shown in other figures. This in turn required smaller system sizes.

\section{Discussion}\label{sec-discuss}
To recapitulate, we have studied periodically driven Anderson insulators with both local and global monochromatic driving starting from the ground state. Our results extend beyond the linear response regime studied by Mott to include both long times and strong driving. We have discussed these results in the setting of a ``phase diagram'' in the frequency and amplitude plane with four distinct regimes. One of these is the traditional linear response regime and the other three: the perturbative non-linear regime, the adiabatic non-linear regime and the (non-linear) enhanced dissipation regime are new and exhibit FLTS with interesting properties. We have presented a 
framework involving pre-existing Mott resonances, field induced Landau-Zener crossings and considerations of adiabaticity and lack thereof to identify these regimes. We have presented several diagnostic quantities that are able to tease these regimes apart. The time-dependence of the energy absorbed distinguishes linear response from all the others, the change of single particle occupations distinguishes linear response and perturbative non-linear response from each other and from the remaining two regimes, the structure of the evolution operator allows us to tell apart the adiabatic non-linear regime from the enhanced dissipation regime. Finally, the spatial inhomogeneity of the 
excitations decreases monotonically as we go away from the linear response regime.

We have provided results from numerical simulations to illustrate the breakdown of linear response and characterize the Floquet regimes. In particular, we have identified the boundary of the linear response regime from numerics. We have shown that the main contributions to  beyond linear response no longer come from resonant pairs, but instead from multiphoton processes and Landau-Zener avoided energy level crossings.
In the simulations we have seen that systems heat up in an active manner when driven beyond linear response and a significant fraction of avoided energy level crossings are traversed near-adiabatically. We have also seen that the dissipative and reactive contributions to  predicted by theory match the data from simulations well in terms of the dependence on field strength.

In closing we note that the defining characteristic of ``plain vanilla'' Anderson insulators, and localized insulators more generally, is a vanishing linear DC conductivity. That by itself does not tell us very much about the state and indeed cannot be distinguished from the conductivity of a band or Mott insulator. To probe deeper into the state it is necessary to move away from this limit. Mott's classic work showed the linear finite frequency response was a probe of two site resonances in the spectrum of the system. In the present paper we have shown that investigating the long time and non-linear response teases out more information about the system. In particular the non-linear response probes the {\it creation} of resonances as the system is made by the application of an electric field to move along a set of non-generic potential configurations starting from a generic disorder configuration. Altogether the full frequency and amplitude response yields a much wider window on the dynamics of the Anderson insulator and we look forward to experiments that will take advantage of this possibility. Examining these questions in quasiperiodic single-particle\cite{aubry1980} and many-body localized\cite{IyerQP, KhemaniCPQP} systems with quite different resonance structures is also an interesting direction for future work.

\section*{Acknowledgements}
DTL thanks Dmitry Kovrizhin and Adam Nahum for helpful discussions. VK and SLS thank R. Nandkishore for previous collaboration on related ideas and A. Lazarides and R. Moessner for many discussions on Floquet physics. This work was supported by the Marshall Aid Commemoration Commission and the Rudolf Peierls Centre for Theoretical Physics (DTL), EPSRC Grant No.~EP/N01930X/1 (JTC) and by US Department of Energy grant No. DE-SC0016244 (SLS). VK is supported by the Harvard Society of Fellows and the William F. Milton Fund. 

\appendix
\renewcommand\thefigure{\thesection.\arabic{figure}} 
\section{Single-site drive}\label{singlesite}
It is also interesting to consider a system driven by a local oscillating potential, rather than an electric field that acts globally. A broadly similar treatment applies to the one presented in Sec.~\ref{analysis}, but with some characteristic differences. It is reassuring that our general approach is useful in a second setting. It is also helpful that some differences arise, as correct capture of these provides a test for simulations. In the following we outline the parallels and differences between the two types of drive, considering specifically the tight binding model with Hamiltonian given by
\begin{align}
H_0 &= - \sum_i \lf \lambda c_i^{\dag}c_{i+1} + \text{h.c.}\ri + \sum_i w_i c_i^{\dag}c_i,
\end{align}
where the site potentials $w_i$ are independent random variables uniformly distributed in $\left[-W,W\right]$. 
The model with single-site drive is
\begin{equation}\label{ssd}
H_\text{SSD}\lf t \ri   = H_0 + v\sin\left(\omega t\right)c_d^{\dag}c_d\,.
\end{equation}
We focus on strong disorder: in this case the density of states in energy per site is $\rho \sim 1/W$ and the localization length is $\xi \sim 1/\ln(W/\lambda)$ in units of the lattice spacing.

Suitable dimensionless measures of the drive frequency and strength in this instance are
\begin{equation}
{\Omega} = \hbar \omega /W\quad {\rm and} \quad {\cal V} = v/W\,.
\end{equation}
The minimum energy difference between two eigenstates with spatial separation $x$ between their localization centers is of order $W(\lambda/W)^x \equiv W \exp(-x/\xi)$.  Equating this to the photon energy $\hbar \omega$, the Mott length in units of $\xi$ is again $r_{\rm Mott} = \ln(1/\Omega)$. Equating the same energy difference to the drive amplitude gives $r_L = \ln(1/{\cal V})$, and a treatment of the condition for adiabaticity following Eq.~(\ref{adiabatic}) gives $r_c = \frac{1}{2}\ln(1/\Omega {\cal V})$.

Adapting the discussion of Rabi oscillations of Mott resonances given in Sec.~\ref{secRabi}, a crucial difference for the case of single-site drive is that the matrix element $\gamma$ does not involve $x_{\rm Mott}$; instead we have simply $\gamma \sim v$. The consequence of this in one dimension is that the linear-response energy absorption rate is proportional to $\omega^2$, without the $\ln^2\omega$ factor that is present in the electrical conductivity. This difference is apparent in our simulations (compare Figs.~\ref{fig:TwoColGD} and \ref{fig:TwoColSSD}). For single-site drive we obtain
\begin{equation}
\Delta E(t)\sim \left\{ \begin{array}{lcl}
\hbar v^2 \rho^2 \xi \omega^2 t & \quad & \omega t < \Omega/{\cal V}\\
& & \\
v (\rho \hbar \omega)^2 \xi && \omega t > \Omega/{\cal V}\,.
\end{array}\right.
\end{equation}

The boundary to the linear response regime, defined as the drive strength at which energy absorption saturates on the timescale of the drive period, is ${\cal V} \sim \Omega$. On the weak-drive side of this boundary, the characteristic length scales have the ordering $r_{\rm Mott} < r_c < r_L$ and only Mott resonances contribute to response. On the strong-drive side, the order is reversed and $r_L < r_c < r_{\rm Mott}$. This is a second difference from the case of global drive (compare with Sec.~\ref{sec:pertnlr}).
\begin{figure}
    \begin{center}
	\includegraphics[width=0.4\textwidth]{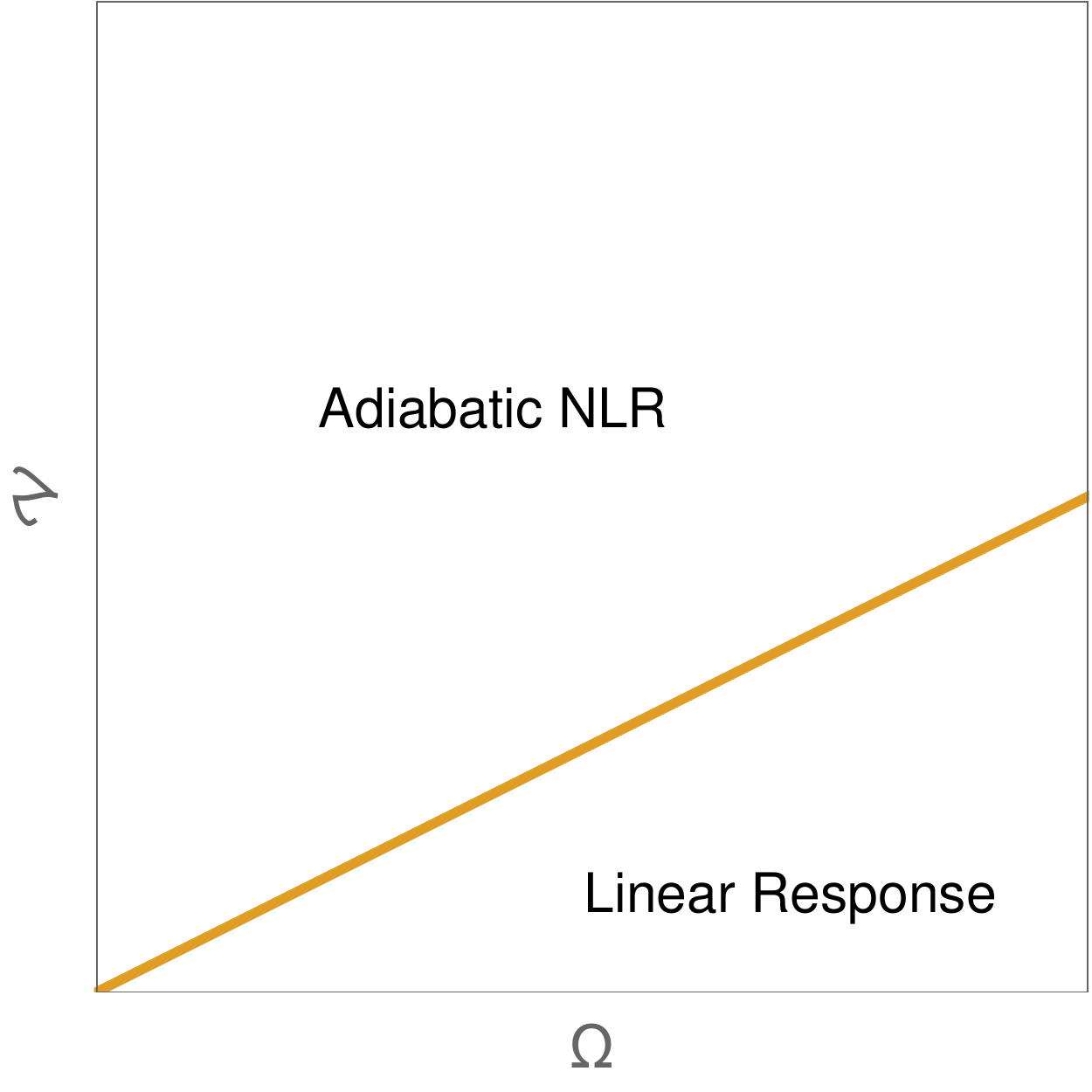}\hspace{2cm}
	\caption{Schematic illustration of regimes of response for an Anderson insulator driven by an oscillating site potential, as a function of the amplitude ${\cal V}$ or potential oscillations and frequency $\Omega$. See main text for distinctions between regimes.}
	\label{fig:regimesSSD}
    \end{center}
\end{figure}

Consider the evolution of eigenstates of the instantaneous Hamiltonian over the course of the drive cycle. An eigenstate that is localized near the drive site may have avoided crossings with other eigenstates that are adiabatic, intermediate or diabatic, according to the spatial separation between the two states 
compared with $\xi r_c$. The average numbers of adiabatic and intermediate crossings are $N_{\rm ad} \sim {\cal V} \xi r_c$ and $N_{\rm int} \sim {\cal V} \xi$ respectively. In the regime we are considering (${\cal V},\, \xi \ll 1$) $N_{\rm int}$ is always small, but there is a change in response at the boundary at which $N_{\rm ad} \sim 1$, implying
\begin{equation}
{\cal V} \sim \frac{1}{\xi \ln(1/\Omega)}\,.
\end{equation}
Above this boundary the system has a large non-linear response, which is principally reactive. Since $N_{\rm int}$ is always small, this model differs from the one with global drive in that it does not have a regime with enhanced dissipation. The regimes of response for the single-site drive are illustrated in Fig.~\ref{fig:regimesSSD}.

\section{Single-site Drive Numerical Simulations}\label{app-numerics}
In this appendix we present results from numerical simulations of a site-disordered, one-dimensional tight-binding model for spinless fermions, driven locally by an oscillating potential at a single site. 
\begin{figure*}
	\includegraphics[width=0.32\textwidth, height=0.27\textwidth]{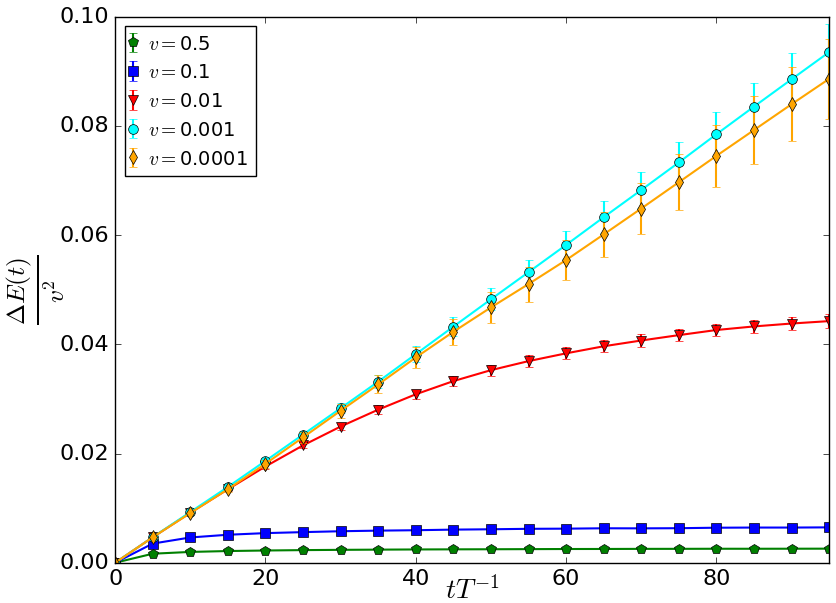}
	\label{fig:AmpSquaredLinearAbsSSD}
	\includegraphics[width=0.33\textwidth, height=0.27\textwidth]{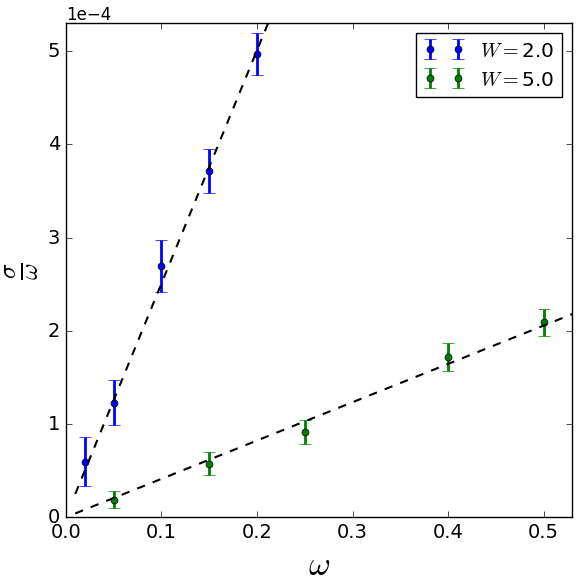}
	\label{fig:MottLawCondSSD}
	\includegraphics[width=0.32\textwidth, height=0.26\textwidth]{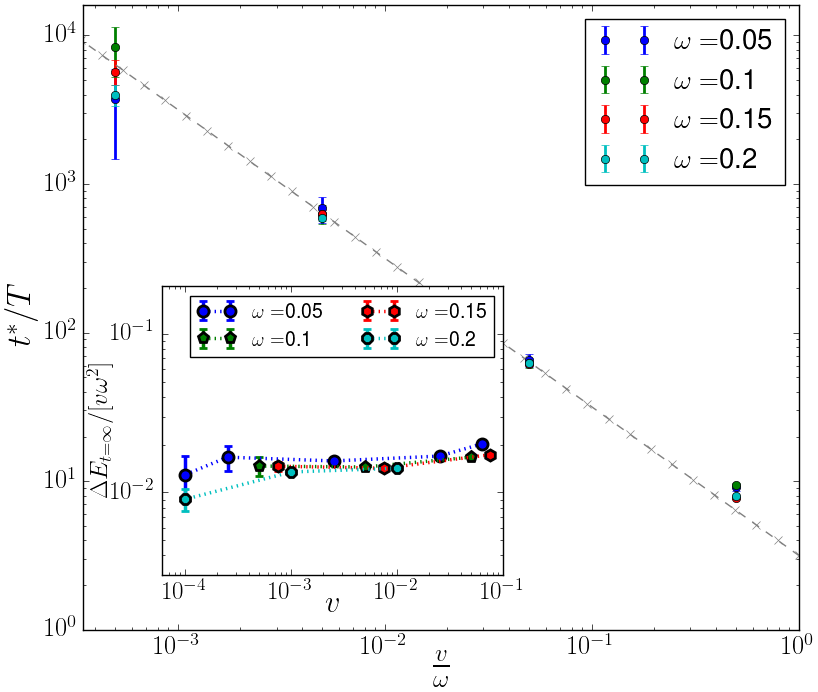}
	\caption{(left) $\Delta E(t)/v^2$ vs $t$ for a range of drive strengths, demonstrating a short-time regime with a constant absorption rate, and saturation at long times. The linear response coefficient  (the value of the `conductivity' $\Delta E(t)/(v^2t)$ at short time) characterizing the absorption rate is independent of drive strength. Points are data measured every five periods, lines are guides to the eye. Parameter values: $\omega = 0.2$ and $N_r \sim 2.5 \times 10^5$. (middle) Frequency dependence of the initial energy absorption rate for weak single-site drive: $\sigma/\omega$ vs $\omega$. As expected from the theory of locally driven Mott resonances, rate $\propto \omega^2$. Parameter values: $v=10^{-4}$ and  $W=2,\,5$. (right) Dependence of saturation time $t^\star$ on drive strength and frequency: $t^*/T$ vs $v/\omega$ on log-log scale. Points: data; dashed line: fit with slope $-1$. (inset) Dependence of $\Delta E(\infty)$ on $\omega$ and drive strength, $v$. Ratios of $\Delta E(\infty)$ to predicted $\omega$-dependence [Eqn.~(\ref{Einftyapp})] vs $\omega$ on log-log scales. Points: data; lines: guides to the eye. (all) Other parameter values: $W=2$, $N_r \sim 5 \times 10^4 \-- 5\times 10^5$.}
	\label{fig:TstarSSD}
	\label{fig:TwoColSSD}
\end{figure*}

We start from the Hamiltonian of Eq.~(\ref{ssd}). The disorder strengths, drive strengths, and drive frequencies are in the same range as in the main text for the global drive. All other considerations (e.g. initial state, boundary conditions, and system size) are as in the main text as well. An overview of the dependence of energy absorbed on time and drive strength is given in Fig.~\ref{fig:TwoColSSD} (left). The variation of the energy absorption rate $\sigma$ with frequency is shown in Fig.~\ref{fig:TwoColSSD} (middle).

It matches the theory of Mott resonances with \emph{single-site} drive and differs from the Mott law for a global drive by a factor of $\ln^2\omega$. Further distinctions between the cases of single-site and global drive are in the dependence of the saturation time $t^*$ and saturation energy $\Delta E(\infty)$ on frequency and drive strength. In detail, the results of App.~\ref{singlesite} give
\begin{equation}\label{t^*app}
\frac{t^*}{T} = \frac{\omega}{2\pi v}
\end{equation}
and
\begin{equation}\label{Einftyapp}
\Delta E(\infty) = \frac{\xi \omega^2 v}{(2W)^2} \,.
\end{equation}
Comparisons consistent with Eq.~(\ref{t^*app}) are shown in Fig.~\ref{fig:TwoColSSD} (right), and with Eq.~(\ref{Einftyapp}) in the inset of Fig.~\ref{fig:TwoColSSD} (right).

\bibliography{bibliography_ordered}

\begin{thebibliography}{52}
\bibitem[{FLT()}]{FLTSnote}
\bibinfo{note}{More generally they can synchronize with multiples of the
  fundamental period.}
\bibitem[{Nom()}]{NominalContributionNote}
\bibinfo{note}{We note the subtlety that if we consider Fourier components of
  the full $\Delta E(t)$ at long times then the zeroth component has a larger
  contribution in some regimes from the nominally reactive (e.g. $\sin^2\omega
  t$ or $|\sin^3 \omega t|$) pieces than from the nominally dissipative pieces.
  However, if we measure $\Delta E(t)$ stroboscopically, then only the
  dissipative piece contributes and all is well.}
\end{thebibliography}
\end{document}